\documentclass{article}
\usepackage{spconf,amsmath,graphicx}
\usepackage{multirow}
\usepackage{graphicx}
\usepackage{amsmath,amssymb}
\usepackage{microtype}
\usepackage{subfigure}
\usepackage{cite}

\usepackage{hyperref}
\usepackage{nicefrac}
\usepackage{booktabs}
\usepackage{makecell}

\usepackage[capitalize,nameinlink]{cleveref}

\newcommand{\bw}{\mathbf{w}}

\newcommand{\bee}{\begin{eqnarray}}
\newcommand{\eee}{\end{eqnarray}}

\newlength{\widebarargwidth}
\newlength{\widebarargheight}
\newlength{\widebarargdepth}

\newcommand{\eat}[1]{}

\newcommand{\btx}{\tilde{\mathbf{x}}}
\newcommand{\bty}{\tilde{y}}

\newcommand{\bx}{\mathbf{x}}

\newcommand{\bz}{\mathbf{z}}

\usepackage{pifont}

\newcommand{\newpara}[1]{\vspace{5pt}\noindent\textbf{#1}}
\usepackage[noend]{algpseudocode}
\usepackage{algorithm}

\title{SUPERVISED ATTENTION FOR SPEAKER RECOGNITION}
\name{Seong Min Kye$^{1}$, Joon Son Chung$^{2}$, Hoirin Kim$^{1}$}
\address{$^1$Korea Advanced Institute of Science and Technology, $^2$Naver Corporation}

\begin{document}
%\ninept
%
\maketitle
\begin{abstract}
The recently proposed self-attentive pooling (SAP) has shown good performance in several speaker recognition systems. In SAP systems, the context vector is trained end-to-end together with the feature extractor, where the role of context vector is to select the most discriminative frames for speaker recognition. However, the SAP underperforms compared to the temporal average pooling (TAP) baseline in some settings, which implies that the attention is not learnt effectively in end-to-end training. To tackle this problem, we introduce strategies for training the attention mechanism in a supervised manner, which learns the context vector using classified samples. With our proposed methods, context vector can be boosted to select the most informative frames. We show that our method outperforms existing methods in various experimental settings including short utterance speaker recognition, and achieves competitive performance over the existing baselines on the VoxCeleb datasets.
\end{abstract}

\section{Introduction}
\label{sec:intro}

Speaker recognition is the process of automatically recognising who is speaking by using the speaker-specific information included in speech waveforms. As the use of voice commands become ubiquitous, automatic speaker verification is an essential security measure to protect the users' security and privacy.
In speaker recognition and verification, a key challenge is to aggregate variable-length input speech into a fixed dimensional vector, which is called an utterance-level representation. In practical scenarios, recording environments can be noisy and parts of the speech may not contain discriminative information of speaker identity. 
% Due to these issues, the methods to aggregate frame-level representations into an utterance-level embedding is a key research item in speaker recognition.

Before the advent of deep neural networks (DNN), i-vector systems with probabilistic linear discriminant analysis (PLDA) have held the state-of-the-art in speaker recognition~\cite{kenny2010bayesian,burget2011discriminatively,matvejka2011full}. However with the advances in deep learning, DNN-based speaker recognition systems have achieved superior performance compared to the i-vector systems~\cite{dehak2010front, garcia2011analysis,prince2007probabilistic}. In recent DNN-based speaker recognition systems, there have been many attempts to extract informative speaker embedding effectively. The na\"ive aggregation method is temporal average pooling (TAP), which represents a simple average pooling along the time axis. However, our voice changes from time to time and also contains short pauses even within utterances. To address this problem, Cai \textit{et al.}~\cite{SAP} proposed self-attentive pooling (SAP) to select informative frames more effectively. In SAP, frame-level features are weighted according to their similarity to the context vector which is a learnable vector. However, SAP often shows lower performance than TAP depending on the training settings, which suggests that the context vector has not been trained effectively to select the most informative frames. 

In order to tackle this problem, we propose methods to train the context vector with {\em explicit} supervision. We propose three variants of the method with classification result over whole training classes. First, we train hidden representation of correctly classified samples and the context vector to have high similarity. Second, contrary to the first method, we train the hidden representation of incorrectly classified samples to be far from the context vector. Lastly, we learn the context vector using both correctly and incorrectly classified samples. These methods allow context vector to capture informative frames which are relevant to the speaker identity.

To show the effectiveness of our proposed methods, we experiment on across various experimental settings. Furthermore, in order to see if the proposed method works well in realistic settings such as short utterance speaker verification, we apply our methods on the state-of-the-art speaker recognition model for short utterances~\cite{kye2020meta}. Since the effectiveness of meta-learning has been demonstrated on this task~\cite{kye2020meta, defence, centroid}, we use our base model as the meta-learning framework proposed in \cite{kye2020meta}.

Our main contributions are as follows: (1) We propose a novel supervised learning method for the context vector in self-attentive pooling (SAP), in which the context vector is optimized with classified samples. (2) Our proposed methods can be implemented in a few lines of code, and only leads to a small increase in the computational cost. (3) To generalize the performance improvement, we experiment in various settings and show consistent improvement over the baselines.
\begin{figure*}[t]
\centering
\includegraphics[width=1\linewidth]{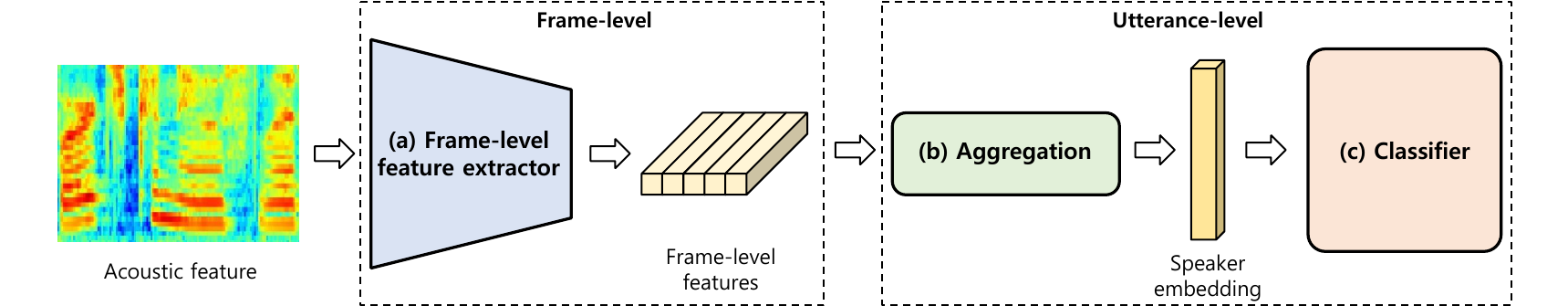}
\caption{\textbf{Overview.} (a) To extract the frame-level features, many speaker recognition systems use frame-level feature extractors such as 1D or 2D CNNs. In this paper, we use the ResNet-34 as the feature extractor which is a type of 2D CNN. (b) In order to represent the speaker into a single fixed vector, we aggregate the frame-level features. In our experiments, the aggregation layer is followed by a fully-connected layer. (c) We finally learn the embedding space to be discriminative using a classifier.}
\label{fig:overview}
\end{figure*}

\section{methods}
\label{sec:pre}

This section describes the baseline aggregation methods, and introduces the proposed supervised attention.

\subsection{$d$-vector based feature extraction}

$d$-vector refers to the general method of extracting speaker embeddings with deep neural networks (DNN). $d$-vector based models have shown to outperform previous i-vector based models on speaker recognition~\cite{Nagrani17,snyder2017deep,snyder2018x,Xie19a}. $d$-vector systems typically contain three key components -- a frame-level feature extractor, a temporal aggregation layer and a training objective function. 
As for the frame-level feature extractor, 1D or 2D convolutional neural networks~\cite{snyder2017deep, snyder2018x, han2020mirnet, kwon2020intra, kye2020meta} and recurrent neural networks~\cite{wang2019centroid, jung2019rawnet} are commonly used. These networks generate frame-level representations from network inputs such as spectrograms or MFCCs. In order to encode the speaker identity in utterance-level representations, various aggregation methods have been proposed, including self-attentive pooling (SAP)~\cite{SAP}, attentive statistic pooling (ASP)~\cite{ASP}, learnable dictionary encoding (LDE)~\cite{SAP} and cross attentive pooling (CAP)~\cite{kye2020cross}. Finally, there are various optimization techniques to train discriminative speaker embeddings. For this purpose, there are various methods, from the na\"ive softmax classifier~\cite{kwon2020intra, TDV, han2020mirnet} to A-Softmax \cite{Asoftmax, SPE, short_jung}, AM-Softmax \cite{AMsoftmax}, AAM-Softmax \cite{AMMsoftmax} and the prototypical loss~\cite{snell2017prototypical, kye2020meta, defence, wang2019centroid}. In this paper, we mainly deal with combination of the prototypical loss and the softmax loss as in \cite{kye2020meta}.

\subsection{Baseline}
This section describes the self-attentive pooling (SAP) introduced in \cite{SAP}. In SAP, frame level representation $\{x_1,x_2, \dots x_L\}$ are fed into non-linear projection network $g_\phi$, which has single fully-connected layer and $\tanh$ non-linear function in order to get hidden representation $\{h_1, h_2, \dots h_L\}$.
\begin{equation}
h_t=g_\phi(x_t)=\tanh(Wx_t + b)
\end{equation}
These hidden representations are used to measure how informative the frames are. Specifically, the dot product of hidden representation $h_t$ and the learnable context vector $\mu$ is used to get attention weight $w_t$.
\begin{equation}
w_t = \frac{\exp(h_t^T\mu)}
{\sum_{t=1}^{T} \exp(h_t^T\mu)}
\label{eq:attention_weight}
\end{equation}
The context vector $\mu$ is used as a representation of informative frames for speaker recognition. It is jointly learned during training without explicit constraint. The aggregated utterance-level representation $e$ is formulated as follows:
\begin{equation}
e = \sum_{n=1}^{T} w_t x_t
\end{equation}

However, training method for context vector raises a new question, which motivates the contributions of this paper -- {\em is this joint training method for context vector sufficient to select the most informative frames?} 
\begin{figure*}
    % \vspace{-0.25in}
\centering
    \subfigure[APF layer]{\includegraphics[width=0.32\linewidth]{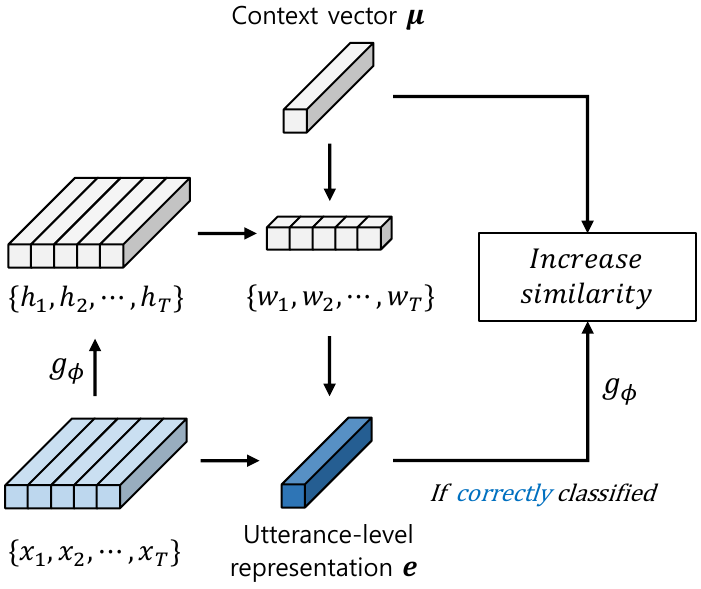}\label{fig:APF}}
    \hspace{0.03in}
    \subfigure[ANF layer]{\includegraphics[width=0.32\linewidth]{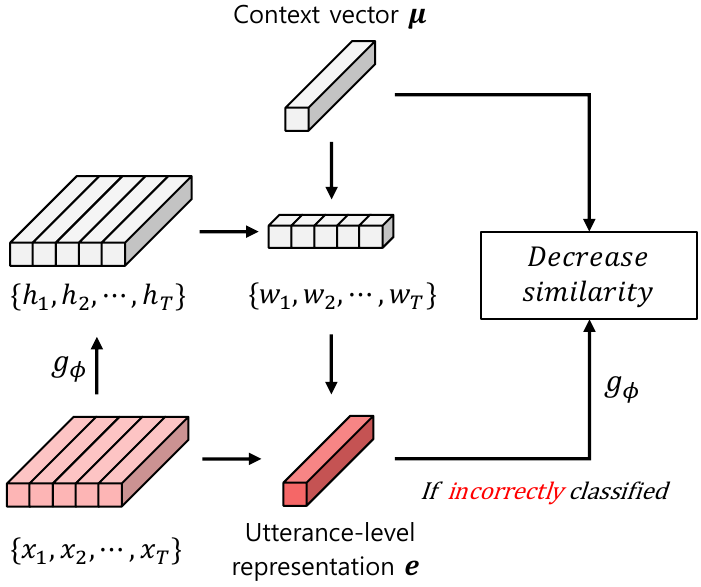}\label{fig:ANF}}
    \hspace{0.03in}
    \subfigure[ADF layer]{\includegraphics[width=0.32\linewidth]{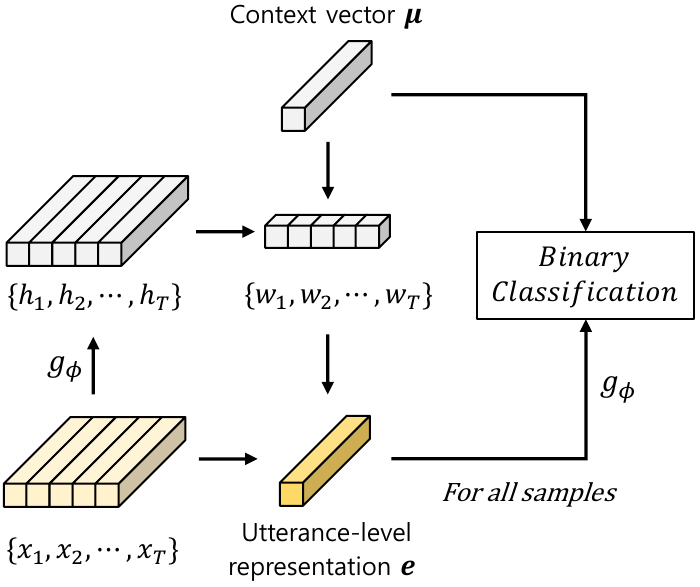}\label{fig:ADF}}
    \caption{Comparison between the three proposed attention methods.}\label{fig:scale_histogram}
    % \vspace{-0.10in}
\end{figure*}

% \begin{figure*}
%     % \vspace{-0.25in}
% \centering
%     \subfigure[APF layer]{\includegraphics[width=0.47\linewidth]{pdfs/AWC_.pdf}\label{fig:AWC}}
%     \hspace{0.3in}
%     \subfigure[ANF layer]{\includegraphics[width=0.47\linewidth]{pdfs/AWE.pdf}\label{fig:AWE}}
%     \caption{Comparison between the two proposed training methods.}\label{fig:scale_histogram}
%     % \vspace{-0.10in}
% \end{figure*}

\subsection{Supervised attentive pooling}
\label{sec:method}

In order to answer this question, we propose a method for training the context vector in a supervised manner. In SAP, context vector is used to screen informative frame-level representation. However, there is no explicit constraint to force the informative frames to get high similarity with the context vector. To overcome this problem, we train the context vector with additional supervision.

\subsubsection{Attention with positive feedback} In SAP, the context vector is trained in an end-to-end manner, and we expect the context vector to be able to select the most informative frames. To enhance the ability to select the most discriminative frames, we first introduce an attention with positive feedback (APF) which uses correctly classified samples $D_{cor}$ for the context vector. We assume that aggregated vector of the correctly classified samples has abundant information about the speaker identity. Therefore, we train the correctly classified samples to be more similar to the context vector.

As shown in Equation~\ref{eq:attention_weight}, the similarity between the hidden representation $h$ and the context vector $\mu$ is measured as the following:
\begin{equation}
h^T \mu = |h||\mu| \cos(h, \mu)
\end{equation}
where $\cos$ denotes the cosine similarity. So, in order to increase the similarity between hidden representations of correctly classified samples and the context vector, we deduct average cosine similarity from the final objective. Here, the reason we feed utterance-level representation $e$ into non-linear projection $g_{\phi}$ is to match the space with the context vector $\mu$.
\begin{equation}
L_{\mu} = \frac{1}{|D_{cor}|}\sum_{e \in D_{cor}} -\cos(g_{\phi}(e),\mu)
\end{equation}

\subsubsection{Attention with negative feedback}
Secondly, we propose the method that reduces the cosine similarity with incorrectly classified samples $D_{mis}$, which we call attention with negative feedback (ANF). To train the context vector, a straight-forward option would be using the correctly classified samples. However, the correctly classified samples constitute the vast majority of the examples during training. In particular, we notice that the training accuracy go up to nearly 100\% when training the ResNet-34 network on the VoxCeleb dataset. Therefore, if we train the context vector with correctly classified samples by increasing similarity with the context vector, nearly every representation would be forced to be similar to the context vectors, making it difficult to find non-informative representations. Moreover, given that we choose the same architecture as \cite{SAP}, where a shallow network is used for attention mechanism, it is much harder to train the context vector to capture informative representations if we use correctly classified samples. Hence, we train the context vector to discriminate non-informative frames as the similarity to incorrectly classified samples $D_{mis}$ becomes low.
\begin{equation}
L_{\mu} = \frac{1}{|D_{in}|}\sum_{e \in D_{in}} \cos(g_{\phi}(e),\mu)
\end{equation}

\subsubsection{Attention with dual feedback}
In a realistic scenario, training accuracy can vary depending on the task. For example, training with angular margin loss may yield low training accuracy. Therefore, in order to generalize the supervised attention framework, we lastly propose the attention with dual feedback (ADF), which utilizes both correctly and incorrectly classified samples. Similarly to APF and ANF, our goal is to make the context vector closer to the correctly classified samples and far from the incorrectly classified samples. Toward this goal, we conduct binary classification using context vector. Specifically, we augment minus context vector, and compose weights of the classifier:
\begin{align}
\omega = \{\bw_{cor}=\mu, \bw_{in}=-\mu\}
\label{eq:binary_classifier}
\end{align}
Then, we classify both correctly and incorrectly classified samples, using classification results (or feedbacks) as their labels. Hence, we can calculate the probability of sample belonging to each weight.
\begin{equation}
p(\bz|e;\theta,\phi,\mu) =
\frac{\exp(g_{\phi}^T(e){\bw_{\bz}})}
{\exp(g_{\phi}^T(e){\bw_{cor}})+\exp(g_{\phi}^T(e){\bw_{in}})}
\label{eq:pixel_pred}
\end{equation}
where $\bz \in \{cor, in\}$ is classification result over whole training classes. With this probability, we use cross-entropy loss for the context vector as following:
\begin{equation}
L_{\mu} = \frac{1}{|D|}\sum_{e \in D} -\log p(\bz|e;\theta,\phi,\mu)
\label{eq:global_loss}
\end{equation}
Here, $D$ is the mini-batch. This loss is simply added to final objective. As a result, this binary classifier allows us to use whole samples in each batch and obtain an appropriate context vector considering both correctly and incorrectly samples.
\section{Experiment}
\label{sec:expemriment}

\subsection{Dataset}
We use the VoxCeleb datasets~\cite{Nagrani17,Chung18a} in our experiments, which are popular text-independent speaker recognition datasets. VoxCeleb1 and VoxCeleb2 contain 1,251 and 5,994 speakers, respectively. The two datasets are mutually exclusive. 

\subsection{Evaluation}
The verification results are measured by the equal error rate (EER) and the minimum  detection cost function (minDCF or $C_{det}^{min}$ at $P_{target}=0.01$)~\cite{Sadjadi2019nist}. Cosine similarity is used as the distance metric.

\begin{table}[t!]
\centering
\small
\caption{The architecture of the frame-level feature extractor based on 34-layer ResNet. The input size is 40 $\times$ T.}
\vspace{+0.1in}
\begin{tabular}{c | c | c}
\hline
stage & output size & ResNet-34\\
\hline
conv1 & $40 \times T \times 32$ &
7 $\times$ 7, 32, stride 1\\
\hline
block1 & $40 \times T \times 32$ &
$\begin{bmatrix}
3 \times 3, 32\\
3 \times 3, 32
\end{bmatrix}
\times 3$\\
\hline
block2 & $20 \times \nicefrac{T}{2} \times 64$ &
$\begin{bmatrix}
3 \times 3, 64\\
3 \times 3, 64
\end{bmatrix}
\times 4$\\
\hline
block3 & $10 \times \nicefrac{T}{4} \times 128$ &
$\begin{bmatrix}
3 \times 3, 128\\
3 \times 3, 128
\end{bmatrix}
\times 6$\\
\hline
block4 & $5 \times \nicefrac{T}{8} \times 256$ &
$\begin{bmatrix}
3 \times 3, 256\\
3 \times 3, 256
\end{bmatrix}
\times 3$\\
\hline
 \end{tabular}
% }
% \vspace{-0.15in}
 \label{tbl:feature_extractor}
\end{table}

\subsection{Experiment setting}

\newpara{Input representations.}
We use 40-dimensional log mel-filterbank (MFB) as the acoustic features, where frame-length is set to 25 milliseconds. We normalize the features along the time axis. In our experiments, voice activity detection (VAD) and data augmentatation (DA) is not applied to the input. When training the models with the classification-based methods (e.g. Softmax, AM-Softmax), we use an input audio segment cropped to 2 seconds. When we implement the learning method proposed in \cite{kye2020meta}, we use the same experimental settings. Specifically, mini-batch is composed of 1 support example and 2 query examples, where they are sampled from 100 classes. Then, the length of the support set is set to 2 seconds, whereas the length of query set is set to 1 to 2 seconds.

\newpara{Trunk architecture.}
We use the ResNet-34 as the frame-level feature extractor. The residual networks are widely used in speaker recognition systems~\cite{SAP,ASP,chung2020delving,huh2020augmentation,defence}. As shown in Table~\ref{tbl:feature_extractor}, we set the number of channels in each residual block to 32-64-128-256. The aggregation layer is followed by a single fully-connected layer with the hidden size of 256. 

\begin{table*}[t!]
\normalsize
\centering
\small
% \vspace{-0.10in}
    \caption{Comparison with the state-of-the-art speaker verification models on full utterance. PL: Prototypical loss; TDV: Time distributed voting; D: Development set; T: Test set; Every model is trained on the VoxCeleb2 dataset~\cite{Chung18a}.}
    \vspace{+0.1in}
 \begin{tabular}{c c c c c | c c }
 \hline
 \multirow{2}{*}{Feature extractor} & \multirow{2}{*}{Feature} & \multirow{2}{*}{Aggregation} & \multirow{2}{*}{Objective} &
 Train &
 \multirow{2}{*}{$C_{det}^{min}$}
 & EER\% \\
   & & & & dataset & & full\\
 \hline
 \hline
  i-vector~\cite{Nagrani17} & - & Supervector & -
  & VoxCeleb1 &
  0.73 & 8.8 \\
  VGG-M~\cite{Nagrani17} & Spectrogram-512 & TAP & 
  Contrastive & VoxCeleb1 &
  0.71 & 7.8 \\
  ResNet-34~\cite{SAP} & MFB-64 & SAP & A-Softmax
  & VoxCeleb1 &
  0.622 & 4.40 \\
  ResNet-34~\cite{SPE} & MFB-64 & SPE & A-Softmax
  & VoxCeleb1 &
  0.402 & 4.03 \\
  TDNN~\cite{ASP} & MFCC-40 & ASP & A-Softmax
  & VoxCeleb1 &
  0.406 & 3.85 \\ 
  ResNet-34~\cite{ASP} & MFB-40 & TAP & PL + Softmax
  & VoxCeleb1 &
  0.418 & 3.81 \\ 
%   ResNet-34~\cite{short_jung} & MFB-64 & LDE w/ FPM-TC & A-Softmax
%   & VoxCeleb1 &
%   \textbf{0.350} & 3.22 \\
  ResNet-34  & MFB-40 & SAP & PL + Softmax
  & VoxCeleb1 &
  0.399 & 3.56 \\
  ResNet-34 (\textbf{Ours}) & MFB-40 & APF & PL + Softmax
  & VoxCeleb1 &
  0.388 & 3.65 \\
  ResNet-34 (\textbf{Ours}) & MFB-40 & ANF & PL + Softmax
  & VoxCeleb1 &
  \textbf{0.380} & \textbf{3.13} \\
  ResNet-34 (\textbf{Ours}) & MFB-40 & ADF & PL + Softmax
  & VoxCeleb1 &
  0.419 & 3.55 \\ 
  \hline
  UtterIdNet~\cite{TDV} & Spectrogram-257 & TDV & Softmax
  & VoxCeleb2 &
  - & 4.26 \\ 
  Thin ResNet-34~\cite{Xie19a} & Spectrogram-257 & GhostVLAD & Softmax & VoxCeleb2 &
  - & 3.22 \\
  ResNet-50~\cite{EAMS} & Spectrogram-512 & TAP & EAMS & VoxCeleb2 &
  0.278 & 2.94 \\
  ResNet-34~\cite{SPE} & MFB-64 & SPE & A-Softmax & VoxCeleb2 &
  0.245 & 2.61 \\
  ResNet-34~\cite{SPE} & MFCC-30 & Statistic Pooling & Softmax & VoxCeleb1\&2 &
  0.268 & 2.31 \\
  ResNet-34~\cite{kye2020meta} & MFB-40 & TAP & PL + Softmax & VoxCeleb2 &
  0.234 & 2.08 \\
%   ResNet-34~\cite{short_jung} & MFB-64 & LDE w/ FPM-TC & A-Softmax
%   & VoxCeleb2 &
%   \textbf{0.205} & 1.98 \\
  ResNet-34  & MFB-40 & SAP & PL + Softmax & VoxCeleb2 &
  0.233 & 2.05 \\
  ResNet-34 (\textbf{Ours}) & MFB-40 & APF & PL + Softmax & VoxCeleb2 &
  0.253 & 1.92 \\
  ResNet-34 (\textbf{Ours}) & MFB-40 & ANF & PL + Softmax & VoxCeleb2 &
  0.226 & \textbf{1.91} \\
  ResNet-34 (\textbf{Ours}) & MFB-40 & ADF & PL + Softmax & VoxCeleb2 &
  \textbf{0.210} & 1.94 \\
 \hline
 \end{tabular}

 \vspace{-0.10in}
    \label{tbl:full_duration}
\end{table*}
\begin{table}[t!]
\centering
\small
% \vspace{-0.15in}
\caption{Performance comparison on original VoxCeleb1 test set~\cite{Nagrani17} with full utterance. Every model is trained on VoxCeleb1 development set.}
\vspace{+0.1in}
 \begin{tabular}{c | c | c | c}
 \hline
Aggregation & Objective & $C^{min}_{det}$ & EER\% \\
 \hline
TAP & Softmax & 0.483 & 5.11 \\
SAP & Softmax & 0.517 & 5.13 \\
APF & Softmax & 0.481 & 4.88 \\
ANF & Softmax & 0.506 & 4.86 \\
ADF & Softmax & \textbf{0.475} & \textbf{4.76} \\
 \hline
TAP & AM-Softmax & 0.425 & 4.25 \\
SAP & AM-Softmax & 0.384 & 4.00 \\
APF & AM-Softmax & 0.400 & 4.19 \\
ANF & AM-Softmax & \textbf{0.359} & 3.96 \\
ADF & AM-Softmax & 0.369 & \textbf{3.79} \\
\hline
TAP & PL + Softmax & 0.418 & 3.81 \\
SAP & PL + Softmax & 0.399 & 3.56 \\
APF & PL + Softmax & 0.388 & 3.65 \\
ANF & PL + Softmax & \textbf{0.380} & \textbf{3.13} \\
ADF & PL + Softmax & 0.419 & 3.55 \\
\hline
 \end{tabular}
 \vspace{-0.2in}
 \label{tbl:result_vanilla}
\end{table}
\begin{table}[t!]
\normalsize
\centering
\small
% \vspace{+0.1in}
    \caption{Verification performance on short utterances. Every model is trained on the VoxCeleb1 development set~\cite{Nagrani17}.}
    \vspace{+0.1in}
 \begin{tabular}{c | c c c}
 \hline
 Model & EER\% & EER\% & EER\% \\
 (Aggregation) & 1s & 2s & 5s \\
 \hline
 \hline
  ResNet34 (TAP) & 7.53 & 5.39 & 4.03 \\
  ResNet34 (SAP) & 7.27 & 5.07 & 3.69 \\
  ResNet34 (APF) & 7.49 & 5.28 & 3.88 \\
  ResNet34 (ANF) & \textbf{6.95} & \textbf{4.52} & \textbf{3.41} \\
  ResNet34 (ADF) & 7.25 & 5.11 & 3.81 \\
 \hline
 \end{tabular}

%  \vspace{+0.1in}
    \label{tbl:short_vox1}
\end{table}
\begin{table*}[t!]
\normalsize
\centering
\small
\vspace{-0.10in}
    \caption{Comparison with the state-of-the-arts speaker verification models on short utterances. $^{\dagger}$: Drawn from \cite{short_jung}; $^{*}$: Applied data augmentation; PL: Prototypical loss; D: Development set; T: Test set; PL is calculated episodically, whereas the softmax-based losses are calculated for the entire classes in training set.}
    \vspace{+0.1in}
 \begin{tabular}{c c c c c c | c c c }
 \hline
 \multirow{2}{*}{Feature extractor} & \multirow{2}{*}{Objective} & \multirow{2}{*}{Aggregation} & \multirow{2}{*}{Feature} &
  Train & Test & EER\% & EER\% & EER\% \\
   & & & & dataset & dataset & 1s & 2s & 5s \\
 \hline
 \hline
  ResNet34~\cite{SPE}$^{\dagger}$ & A-Softmax & SPE & MFB-64
  & Vox2(D) & Vox1(T) &
  11.12 & 4.93 & 2.98 \\
  ResNet34~\cite{short_jung}$^{\dagger}$ & A-Softmax & LDE w/ FPM-TC & MFB-64
  & Vox2(D) & Vox1(T) &
  5.92 & 3.38 & 2.17 \\
 ResNet34~\cite{kye2020meta} & PL+Softmax & TAP & MFB-40
  & Vox2(D) & Vox1(T) &
  4.77 & 3.00 & 2.20 \\
 ResNet34 & PL+Softmax & SAP & MFB-40
  & Vox2(D) & Vox1(T) &
  4.64 & 2.90 & 2.17 \\
 ResNet34 & PL+Softmax & APF & MFB-40 & Vox2(D) & Vox1(T)
 &  4.70 & 2.91 & \textbf{2.01} \\
 ResNet34 & PL+Softmax & ANF & MFB-40
  & Vox2(D) & Vox1(T) &
  \textbf{4.49} & 2.88 & 2.04 \\
 ResNet34 & PL+Softmax & ADF & MFB-40
  & Vox2(D) & Vox1(T) &
  4.72 & \textbf{2.84} & 2.09 \\
  \hline
 Thin ResNet34~\cite{Xie19a} & Softmax & GhostVLAD & Spec-257
  & Vox2(D) & Vox1(D+T) &
 12.71 & 6.59 & 3.34 \\
 ResNet34~\cite{short2020} & AM-Softmax & SAP & MFB-80
  & Vox2(D)* & Vox1(D+T) & 9.91 & 4.48 & 2.26 \\
 ResNet34~\cite{kye2020meta} & PL+Softmax & TAP & MFB-40 & Vox2(D) & Vox1(D+T) &
  5.31 & 3.15 & 2.17 \\
 ResNet34 & PL+Softmax & SAP & MFB-40 & Vox2(D) & Vox1(D+T) &
 5.34 & 3.15 & 2.20 \\
 ResNet34 & PL+Softmax & APF & MFB-40 & Vox2(D) & Vox1(D+T) &
  5.40 & 3.08 & 2.15 \\
 ResNet34 & PL+Softmax & ANF & MFB-40
  & Vox2(D) & Vox1(D+T) &
  \textbf{5.18} & \textbf{3.00} & \textbf{2.04} \\
 ResNet34 & PL+Softmax & ADF & MFB-40
  & Vox2(D) & Vox1(D+T) &
  5.41 & 3.10 & 2.15 \\
 \hline
 \end{tabular}

 \vspace{-0.15in}
    \label{tbl:short_duration}
\end{table*}

\newpara{Training objective function.}
 In this paper, we implement Softmax, AM-Softmax and prototypical loss functions. Before explaining the objectives, we will define speaker embedding as $\bx$ for clarity.
 
 The softmax loss is calculated with the softmax function followed by the cross-entropy loss. It can be formulated as:
 \begin{equation}
L_s = -\frac{1}{B}\sum_{i=1}^{B} \log \frac{e^{d(\bx_i,w_{y_i})}}{\sum_{j=1}^C e^{d(\bx_i,w_j)}}
\end{equation}
where $w$, $B$ and $d$ are the set of weights for the whole training classes and batch size, and distance metric respectively. We use the same distance metric as \cite{kye2020meta}, where the distance is cosine similarity with scale of input embedding.
\begin{align}
    d(\mathbf{a}_1, \mathbf{a}_2) = \frac{\mathbf{a}_1^T \mathbf{a}_2}{\|\mathbf{a}_2\|_2} =\|\mathbf{a}_1\|_2 \cdot \cos(\mathbf{a}_1, \mathbf{a}_2)
    \label{eq:dist_metric}
\end{align}

AM-Softmax is an advanced version of the softmax loss. This loss gives margin to a decision boundary in order to reduce intra-class variance and increase inter-class variance. The loss is formulated as:
 \begin{equation}
L_{AM} = -\frac{1}{B}\sum_{i=1}^{B} \log \frac{e^{s(\cos(\theta_{i,y_i})-m)}}
{e^{s(\cos(\theta_{i,y_i})-m)} + \sum_{j \neq y_i} e^{s(\cos(\theta_{i,j}))}}
\end{equation}
where scaling $s$ and margin $m$ are set to 40 and 0.1, respectively.

In order to compute the prototypical loss (PL), each mini-batch must be organised into a support set $\mathcal{S} = \{(\bx_i,y_i)\}_{i=1}^{N \times K}$ and a query set $\mathcal{Q} = \{(\btx_i,\bty_i)\}_{i=1}^{N \times M}$ , where $y,\bty \in \{1,\dots,N\}$ are the class labels in the mini-batch. If we define $S_c$ as the support set of class $c$, we can compute the prototype for each class as:
\begin{align}
P_c = \frac{1}{|\mathcal{S}_c|} \sum_{\bx \in \mathcal{S}_c} \bx
\end{align}
With these prototypes, we finally obtain prototypical loss:
 \begin{equation}
L_{PL} = -\frac{1}{|\mathcal{Q}|}\sum_{(\btx,\bty) \in \mathcal{Q}} \log \frac{e^{d(\btx_i,P_{\bty_i})}}{\sum_{j=1}^N e^{d(\btx_i,P_j)}}
\end{equation}
where we use distance metric in Equation~\ref{eq:dist_metric}. In \cite{kye2020meta}, the author proposes prototypical loss with softmax loss for the discriminative embedding space. In this paper, we adopt the same framework for prototypical loss:
\begin{align}
L = L_{PL} + L_s
\end{align}
Unlike Softmax and AM-Softmax, in this setting, each mini-batch is episodically optimized with the prototypical loss~\cite{protonet} and further optimized with the softmax loss. The feedback is generated by the softmax loss.

\newpara{Implementation details.}
We use the SGD optimizer with Nesterov momentum of 0.9, and the initial learning rate and the weight decay are set to 0.1 and 0.0001 respectively. We use the same learning rate scheduler as \cite{kye2020meta, SPE}. Specifically, we decay learning rate by a factor of 10 until convergence. All of our experiments are trained on NVIDIA 2080 Ti GPUs.

\subsection{Comparison with recent methods on full utterance}
In Table~\ref{tbl:full_duration}, we compare our proposed methods with other state-of-the-art models. Our methods are applied to the base model, which combines the prototypical loss and the softmax loss. For the VoxCeleb1 dataset, our method ANF outperforms other baseline models with an EER of 3.13\%. Using the VoxCeleb2 dataset for training, ANF also obtains the superior performance with an EER of 1.91\%. Moreover, all of our supervised attention methods outperform the TAP and SAP in the same setting. This improvement shows that supervision method is helpful for attention mechanism.

\subsection{Comparison with self-attentive pooling}
To show the effectiveness of our models, we train the models with various loss function on the VoxCeleb1 dataset~\cite{Nagrani17}. Table~\ref{tbl:result_vanilla} shows the results using the classification objectives (i.e. Softmax, AM-Softmax) which are based on the single task of classifying the entire training classes, and also the results using the meta-learning framework~\cite{kye2020meta}. We observe that the self-attentive pooling (SAP)~\cite{SAP} underperforms compared to the temporal average pooling (TAP) with the softmax loss. In this setting, SAP does not seem to be able to select informative frames well. Thus, it degrades the performance compared to TAP which averages frames evenly. However, our proposed methods outperform the TAP and SAP in both classification objectives. In our methods, since the context vector is trained in a supervised manner, this constraint enhances the selective ability of the context vector to find the informative frames. The difference between SAP and our methods is only the explicit loss for the context vector. Among our methods, we see that ADF performs best for classification objectives. Furthermore, when we experiment using the meta-learning setting (PL + Softmax), SAP outperforms TAP by 6.56\%. However, APF underperforms SAP with slight margin. With ANF, we can achieve much superior performance compared to the other baselines, outperforming the SAP by 12.07\%.

\subsection{Duration robustness}
We then examine how these pooling methods work for various speech duration. We use base model proposed in \cite{kye2020meta} which is the state-of-the-art speaker recognition model on short utterance. We first experiment with the same settings on VoxCeleb1. When we evaluate the EER, we enroll with the full utterance and test with short cropped utterances. As shown in Table~\ref{tbl:short_vox1}, SAP outperforms TAP in this experiment setting, getting better results for every duration. Unlike in the baseline methods, APF and ADF get achieves weaker performance compared to SAP, but has marginal improvement over TAP. However, ANF achieves a large improvement over the baselines. It shows the best performance for all durations, especially outperforming the TAP by 15.38\% in the 5-second experiment.

To show the effectiveness of our methods on a larger dataset, we train the the model on the VoxCeleb2 development set and test on the VoxCeleb1 dataset. In Table~\ref{tbl:short_duration}, we compare our methods with previous state-of-the-art speaker verification models on short utterance, which are trained on the VoxCeleb2 dataset. The results in the upper rows in Table~\ref{tbl:short_duration} are test on the original VoxCeleb1 test set containing 40 speakers, whereas the results of the lower rows are tested on all speakers in the VoxCeleb1 dataset containing 1,251 speakers. Note that the VoxCeleb1 and VoxCeleb2 datasets are mutually exclusive. In the latter setting, we randomly sample 100 positive pairs and 100 negative pairs for each speaker. If the data duration is less the required length, we simply use the full available segment. 
% All experiments were registered with full utterance. 
% This evaluation setting allows us to get more reliable result than original VoxCeleb1 test set, which is limited to 40 classes. 
We see that our methods outperforms other baselines, resulting the new state-of-the-art performance on short utterance scenarios. 
% As in the VoxCeleb1 , ANF performs well in this training setting.
%  \vspace{-0.15in}

%  \vspace{-0.15in}
\section{Conclusion}
\label{sec:conclusion}
We proposed a novel learning strategy for an attention mechanism that learns context vector in a supervised manner to help the context vector to select more informative frames. To overcome the problem of the existing methods where the context vector of the attention system is learned end-to-end, we learn the context vector by utilizing correctly or incorrectly classified according to the result of the classifier. In other words, we propose several novel extensions of self-attentive pooling, and show significant improvements across various settings and datasets. Further analysis of the duration robustness proves the possibility of using our proposed methods. As future work, it would be beneficial to apply the strategy to attention mechanisms in various fields such as natural language processing and computer vision.

\newpara{Acknowledgements.} This research was supported by the Korean MSIT (Ministry of Science and ICT), under the National Program for Excellence in SW (2016-0-00018), supervised by the IITP (Institute for Information \& communications Technology Planning \& Evaluation)

\clearpage 
\bibliographystyle{IEEEbib}
\bibliography{shortstrings,vgg_local,mybib}

\end{document}